\documentclass[3p,times,twocolumn]{elsarticle}

\makeatletter
\def\ps@pprintTitle{%
 \let\@oddhead\@empty
 \let\@evenhead\@empty
 \def\@oddfoot{}%
 \let\@evenfoot\@oddfoot}
\makeatother

\usepackage{amsmath}

\usepackage{amssymb}

\usepackage[figuresright]{rotating}

\begin{document}

\begin{frontmatter}




\title{Latest results of NEXT-DEMO, the prototype of the NEXT 100 double beta decay experiment}


\author[1]{L. Serra\corref{cor1}}
\author[1]{D. Lorca}
\author[1]{J. Mart\'in-Albo}
\author[1]{M. Sorel}
\author[1]{J.J. G\'omez-Cadenas}
\author{on behalf of the NEXT Collaboration}

\address[1]{Instituto de F\'isica Corpuscular (IFIC), CSIC \& Universidad de Valencia }
\cortext[cor1]{Corresponding author e-mail: luis.serra@ific.uv.es}

\begin{abstract}
NEXT-DEMO is a 1:4.5 scale prototype of the NEXT100 detector, a high-pressure xenon gas TPC that will search for the neutrinoless double beta decay of $^{136}$Xe. X-ray energy depositions produced by the de-excitation of Xenon atoms after the interaction of gamma rays from radioactive sources have been used to characterize the response of the detector obtaining the spatial calibration needed for close-to-optimal energy resolution. Our result, 5.5\% FWHM at 30 keV, extrapolates to 0.6\% FWHM at the Q value of $^{136}$Xe. Additionally, alpha decays from radon have been used to measure several detection properties and parameters of xenon gas such as electron-ion recombination, electron drift velocity, diffusion and primary scintillation light yield. Alpha spectroscopy is also used to quantify the activity of radon inside the detector, a potential source of background for most double beta decay experiments.
\end{abstract}

\begin{keyword}
time projection chamber \sep x-ray \sep alpha decay \sep double beta decay  \sep NEXT


\end{keyword}

\end{frontmatter}


\section{The NEXT-DEMO prototype}
\label{prototype }

NEXT-100 is a 100-kg high-pressure xenon gas electroluminescent TPC \cite{tdr}. It will search for the neutrinoless double beta decay of $^{136}$Xe in the Laboratorio Subterr\'aneo de Canfranc (LSC). The features that make NEXT a powerful $\beta\beta0\nu$ experiment are very good energy resolution, tracking capabilities and scalability to large detector masses.

\begin{figure}[htb]
\label{demo}
\centering
\includegraphics[width=0.26\paperwidth]{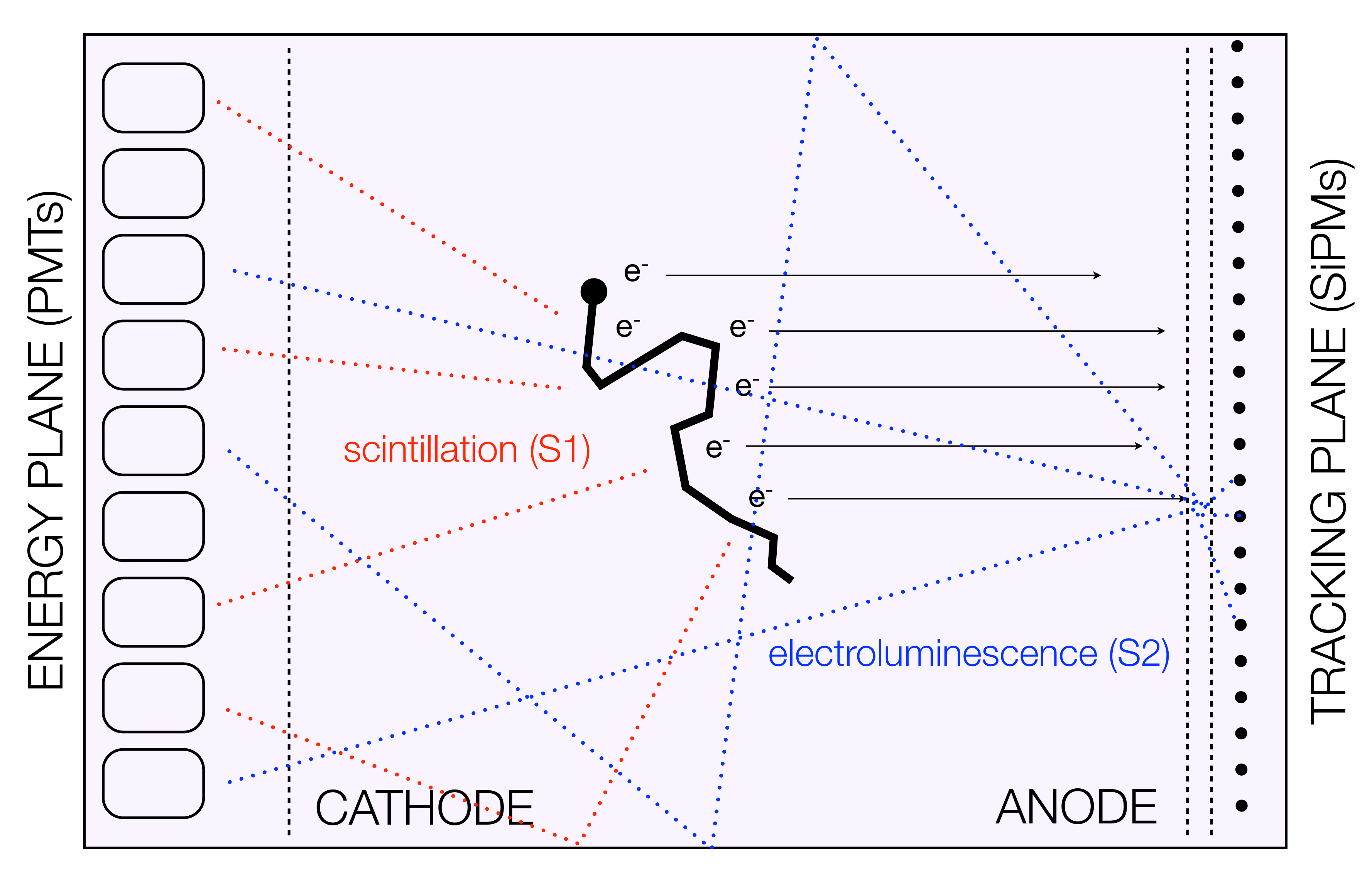}
\caption{ \small The NEXT detector concept. A plane of PMTs located behind a transparent cathode detects both S1 and S2 signal measuring start of event time and the energy of event. On the other side, a plane of SiPMs detects the forward EL light, providing topological information of the event.}
\end{figure}

NEXT-DEMO is a 1-kg prototype built to demonstrate the detector concept to be used in NEXT 100,outlined in Fig.\ \ref{demo}. The xenon active volume of the TPC comprises a 30 cm drift region, operated at a drift voltage between 200-1000 V $\cdot$ cm$^{-1}$ and a 0.5 cm EL region with a reduced electric field of 1-2 kV $\cdot$ cm $^{-1}$ $\cdot$ bar$^{-1}$. The pressure is 10 bar for all the studies presented here. 

It has been running for 3 years using different radioactive sources. Some measurements made so far include energy resolution, imaging of single and double electron tracks and xenon gas properties (drift velocity, diffusion) \cite{initial, er, sipm, Lorca, alphas}.

\section{Studies with electromagnetic depositions}

X-ray depositions have been used to determine the energy response of the TPC, needed to achieve the goal energy resolution \cite{Lorca}. XY energy response has been measured in the active volume of the detector in order to homogenize the response in the energy plane.

\begin{figure}[htb]
\label{fig:dv}
\centering
\includegraphics[width=0.32\paperwidth]{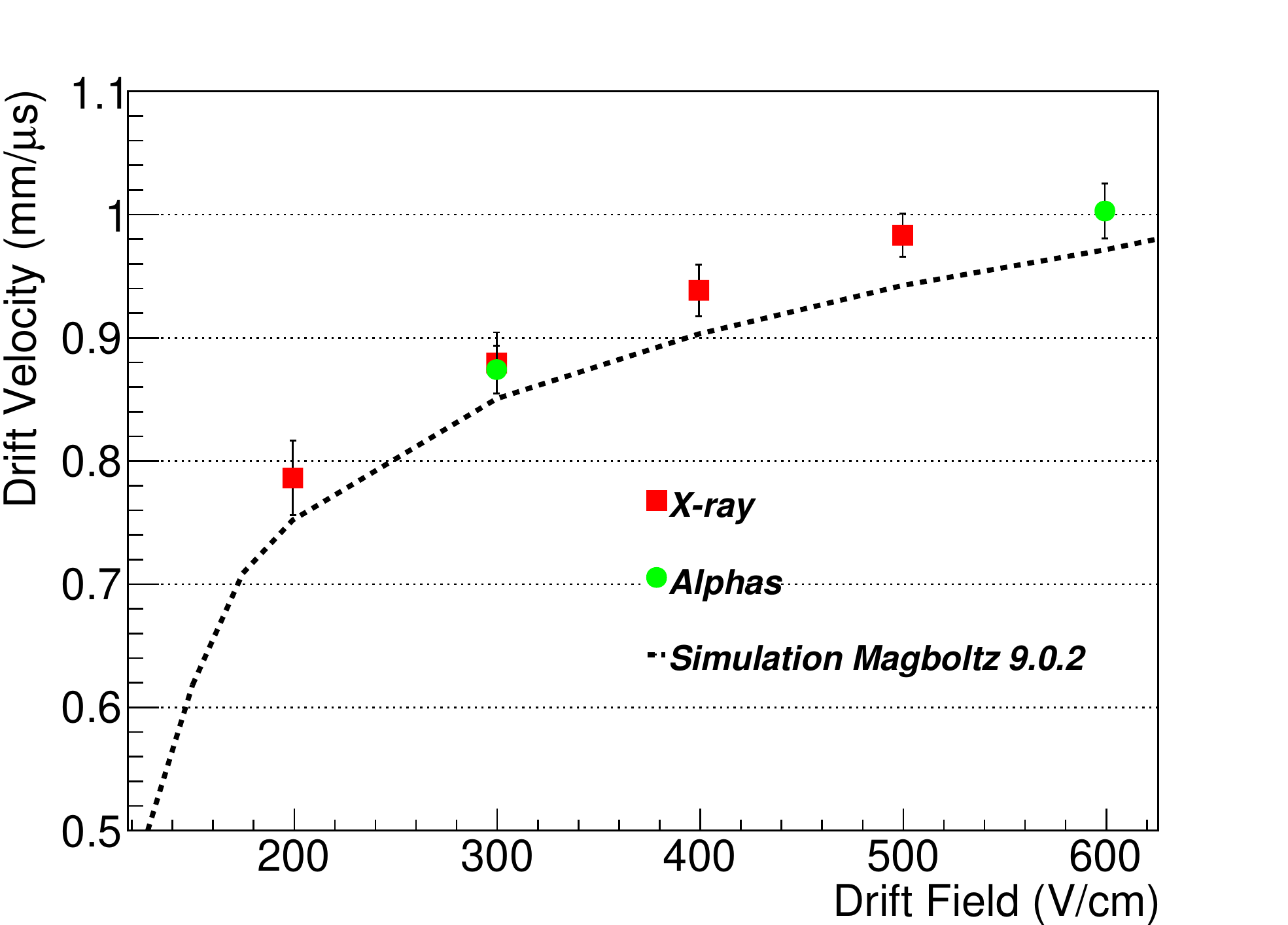}
\caption{ \small Electron drift velocity in NEXT-DEMO for X-ray and alpha decays compared to simulations with Magboltz for pure xenon. \cite{magboltz}}
\end{figure}

Drift velocity is obtained by studying the temporal distribution of these events, shown in Fig.\ \ref{fig:dv}. Using the primary scintillation we are able to measure the drift time of the electrons coming from the cathode, the full drift length, and obtain the drift velocity. Results are consistent with previous measurement using alpha particle depositions \cite{alphas}. In addition, these events are used to monitor the gas quality of the detector, homogenizing the response of different data runs. 

\begin{figure}[htb]
\label{fig:Na22}
\centering
\includegraphics[width=0.32\paperwidth]{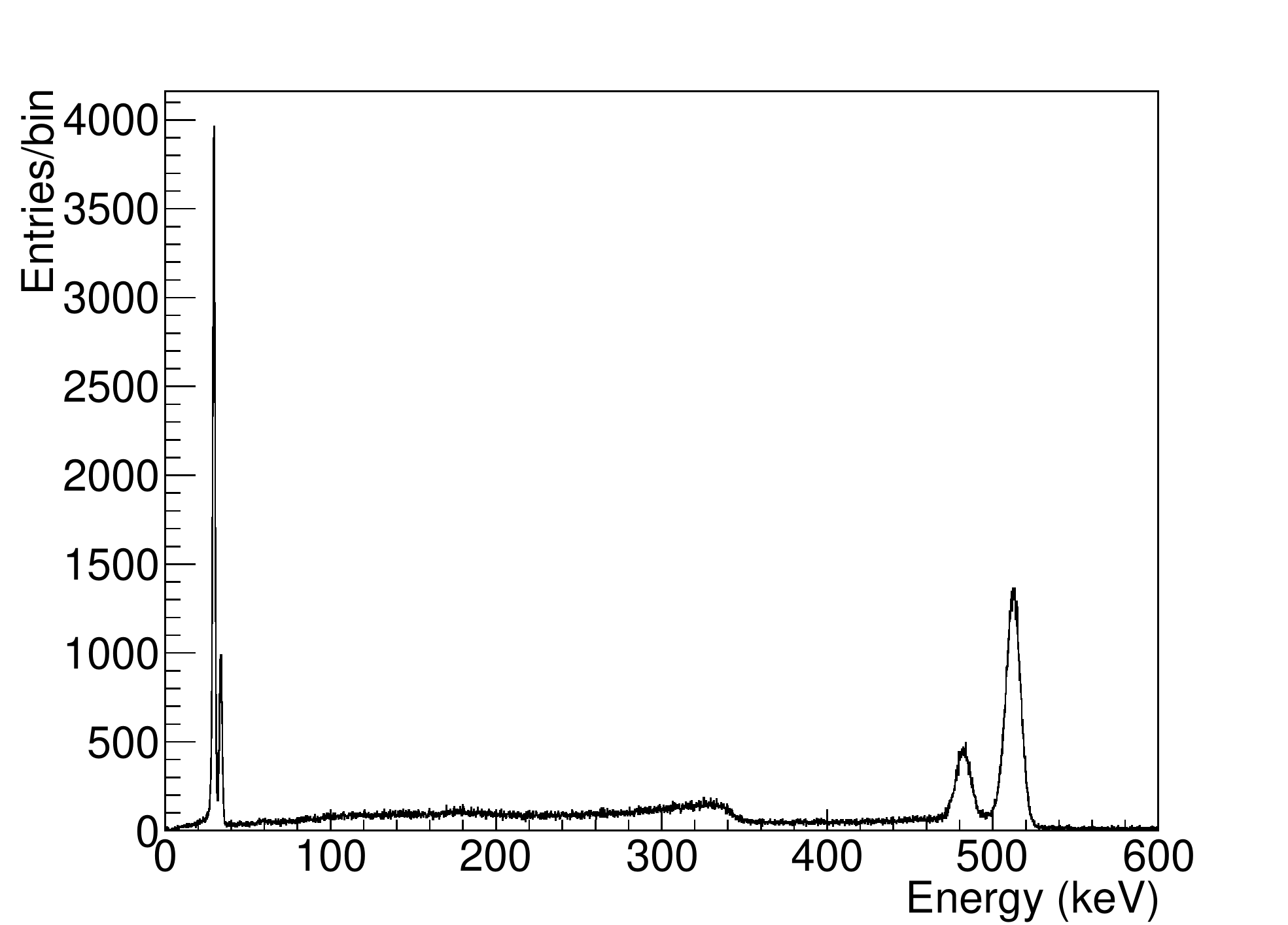}
\caption{ \small Energy spectrum of $^{22}$Na measured in NEXT-DEMO.\ The energy resolution achieved was 0.62$\%FWHM$ and 0.73$\%FWHM$ at $Q_{\beta\beta}$ for the X-ray and photoelectric peaks, respectively. Resolution is scaled as $1/\sqrt{E}$}.
\end{figure}

Energy released by gamma particles coming from $^{22}$Na and $^{137}$Cs sources have been corrected in drift direction, due to attachment, and XY according to individual PMT maps. The corrected energy of the event is calculated by the weighted sum of the contributions of each individual PMT. This method is used to calibrate the energy in all the energy range. Once the energy spectrum is corrected, energy resolution for the X-ray peak as well as the photoelectric peak are obtained. 

\begin{table}[h]\scriptsize
\label{er}
\begin{tabular}{cccc}
\hline \\
\textbf{Source} & \textbf{Energy (keV)} & \textbf{\begin{tabular}[c]{@{}c@{}}Energy Resolution\\ \%FWHM\end{tabular}} & \textbf{\begin{tabular}[c]{@{}c@{}}Energy Resolution\\ \%FWHM@Qbb\end{tabular}} \\
\hline \\
\textbf{$^{137}$Cs}  & \textbf{30}           & \textbf{5.54}                                                               & \textbf{0.61}                                                                   \\
\textbf{$^{137}$Cs}  & \textbf{661.7}        & \textbf{1.58}                                                               & \textbf{0.81}                                                                   \\
\textbf{$^{22}$Na}   & \textbf{30}           & \textbf{5.69}                                                               & \textbf{0.62}                                                                   \\
\textbf{$^{22}$Na}   & \textbf{511}          & \textbf{1.62}                                                               & \textbf{0.73}        \\
\hline \\                                                          
\end{tabular}
\caption{\small Energy resolution for the X-ray and photo-electric peaks of $^{137}$Cs and $^{22}$Na. Resolution better than 1\% FWHM is achieved for Q$_{\beta\beta}$ in all the peaks. }
\end{table}

These results match the goal for the NEXT-100 detector, as they are below 1\% at Qbb of $^{136}$Xe,  as seen in Table 1.

\section{Studies with Alpha particle energy depositions}

Alpha particle energy depositions can be used to study gas Xenon properties as well as detector performance \cite{alphas}. Radon is a potential background for double beta decay experiments. Understanding the control of radon is key for low background experiments like NEXT.  	

In these studies we have measured radon alpha decays coming from a $^{226}$Ra source connected to the gas system of the detector with different drift voltage configurations. 

\begin{figure}[htb]
\centering
\includegraphics[width=0.27\paperwidth]{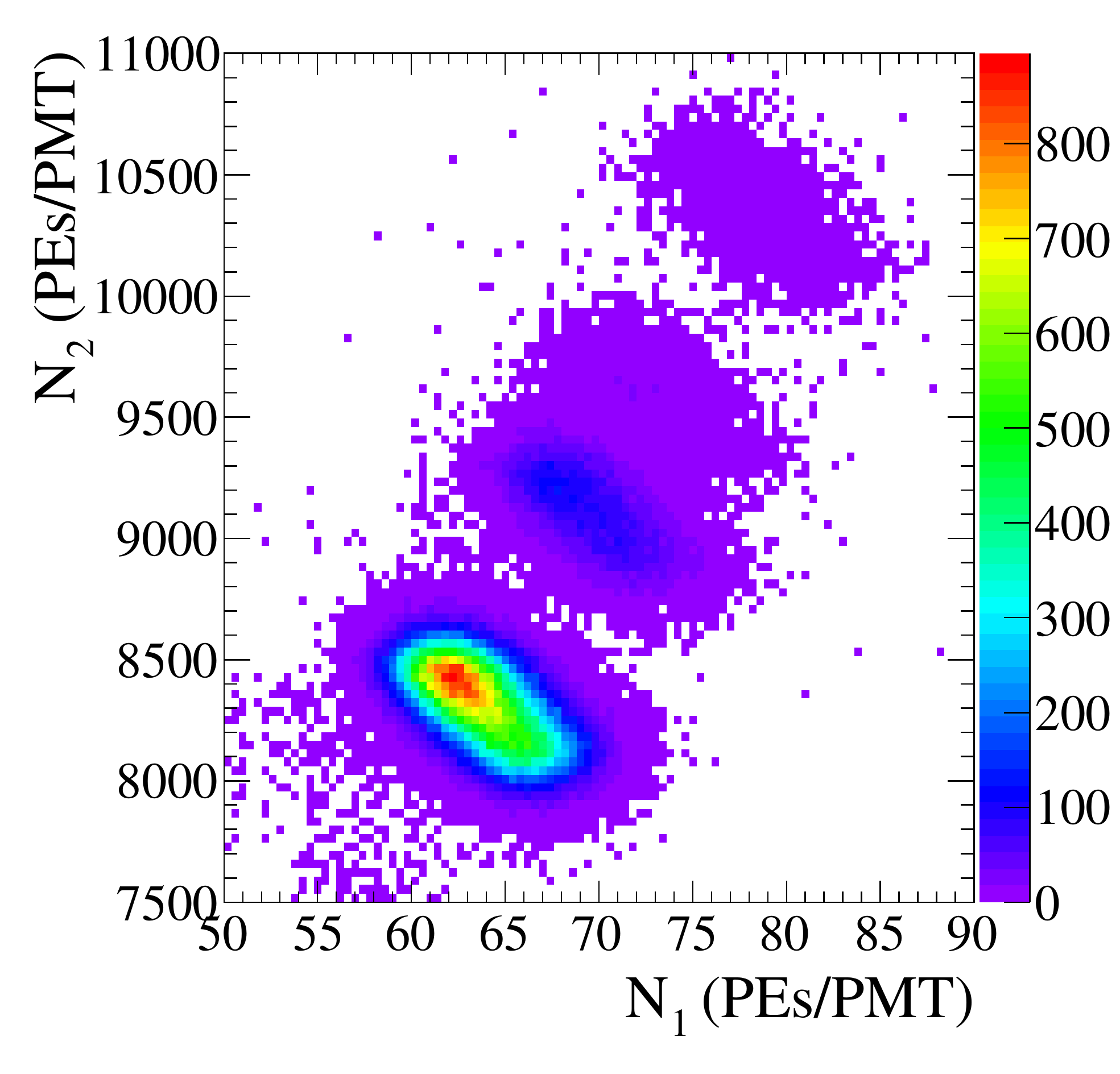}
\caption{ \small Ionization signal (N$_{2}$) versus primary scintillation signal (N$_{1}$) for drift field of 1 kV/cm. Four alpha decays can be seen, from lower to higher energies: $^{222}$Rn, $^{218}$Po, $^{220}$Rn, $^{216}$Po. }
\end{figure}

The energy deposited in the chamber is divided in scintillation (S1) and ionization (S2) of the atoms in the gas. Because of the big amount of electrons generated by the alpha particle some of the free electrons recombine and some energy moves from S2 to S1 signal. Because the electrons lost by recombination transform into scintillation light, there is an anti-correlation between those two signals. Thus we find an anti-correlation between ionization and scintillation signals that can be used to improve the energy resolution \cite{alphas2}. This effect grows when the drift field decreases. 

Using the anti-correlation we have defined the energy as the sum of the S1 signal plus the S2 signal weighted by the optical gain of our detector where $\lambda$ is a calibration constant and $\eta_{EL}$ is the optical gain of the detector: 

\begin{equation}
E = \lambda  (N_{1} + \frac{N_{2}}{\eta_{EL}})
\end{equation}

\begin{figure}[htb]
\label{energy}
\centering
\includegraphics[width=0.32\paperwidth]{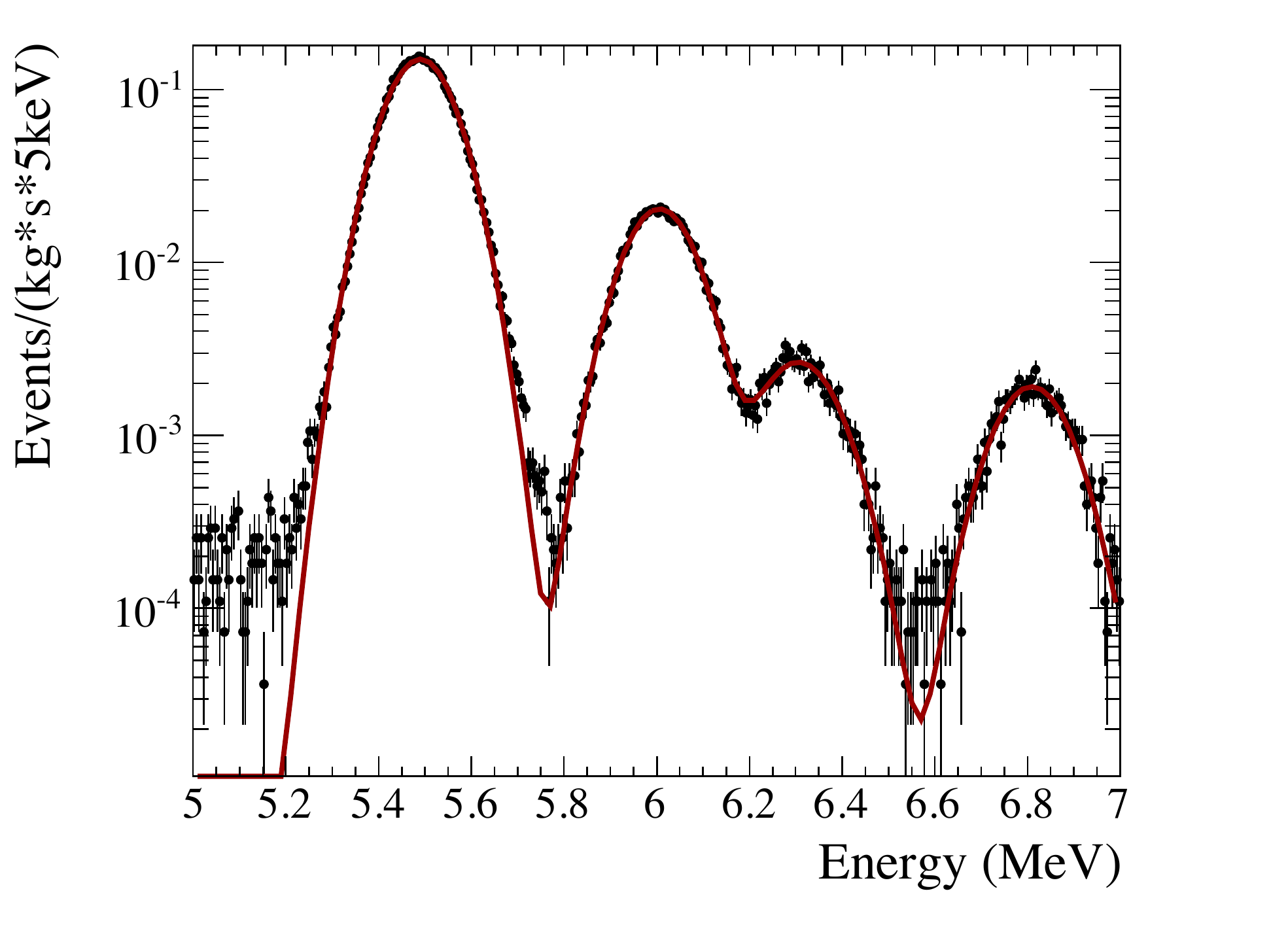}
\caption{ \small Energy spectrum for alpha candidate events in NEXT-DEMO for the 1 kV/cm drift field run. Plot is in log scale. A fit to the data is also shown, as a sum of four Gaussian functions. }
\end{figure}

The two peaks with lower energy (and higher intensity) from Fig.\ 5 correspond to the alpha decays of $^{222}$Rn (5.49 MeV) and $^{218}$Po (6.00 MeV) from the alpha source, while the other two correspond to $^{220}$Rn (6.29 MeV) and $^{216}$Po (6.78 MeV) from natural impurities in the xenon gas. We have obtained an energy resolution in the peak of 5.49 MeV of 2.83\%FWHM, compared to 8\% FWHM of a previous study in NEXT-DEMO. 

We were also able to identify another natural chain of radon that was present in the gas probably coming from natural radon inside the gas.

\section{Conclusions}

We reported on X-ray point-like depositions to measure the energy response of the NEXT-DEMO detector. An energy resolution better than 1\% FWHM at Q$_{\beta\beta}$ of $^{136}$Xe, goal of the NEXT-100 detector, has been obtained for two different radioactive sources, $^{137}$Cs and $^{22}$Na. In addition alpha decays from $^{226}$Ra have been used to study the properties of $^{222}$Rn inside the chamber. Electron-ion recombination has been measured and exploited to improve the energy resolution, which is limited by the low statistics of the S1 signal. Four alpha decays have been observed providing information on the presence of different decay lines inside the detector. 

\section{Acknowledgement}

The authors would like to acknowledge the support of
the Ministerio de Economa y Competitividad of Spain
under grants CONSOLIDER-Ingenio 2010 CSD2008-
0037 (CUP), FPA2009-13697-C04-04 and FIS2012-
37947-C04. Also the support of the European Research
council advanced grant 339787 - NEXT.




\nocite{*}
\bibliographystyle{elsarticle-num}
\bibliography{martin}



\end{document}